\documentclass{PoS}

\title{Structure of a collisionless pair jet in a magnetized electron-proton plasma: Flow-aligned magnetic field}

\ShortTitle{Collisionless electron-positron jet}

\author{\speaker{Mark Eric Dieckmann}\thanks{A footnote may follow.}\\
        ITN, Link\"oping University, 60174 Norrk\"oping, Sweden\\
        E-mail: \email{mark.e.dieckmann@liu.se}}

\author{Martin Falk\\
        ITN, Link\"oping University, 60174 Norrk\"oping, Sweden\\
        E-mail: \email{martin.falk@liu.se}}

\author{Peter Steneteg\\
        ITN, Link\"oping University, 60174 Norrk\"oping, Sweden\\
        E-mail: \email{peter.steneteg@liu.se}}
        
\author{Doris Folini\\
        CRAL, \'Ecole Normale Sup\'erieure, 69622 Lyon, France\\
        E-mail: \email{doris.folini@ens-lyon.fr}}
        
\author{Ingrid Hotz\\
        ITN, Link\"oping University, 60174 Norrk\"oping, Sweden\\
        E-mail: \email{ingrid.hotz@liu.se}}
        
\author{Aida Nordman\\
        ITN, Link\"oping University, 60174 Norrk\"oping, Sweden\\
        E-mail: \email{aida.vitoria@liu.se}}
        
\author{Pierangelo Dell'Acqua\\
        ITN, Link\"oping University, 60174 Norrk\"oping, Sweden\\
        E-mail: \email{pierangelo,dellacqua@liu.se}}
        
\author{Anders Ynnerman\\
        ITN, Link\"oping University, 60174 Norrk\"oping, Sweden\\
        E-mail: \email{anders.ynnerman@liu.se}}
        
\author{Rolf Walder\\
        CRAL, \'Ecole Normale Sup\'erieure, 69622 Lyon, France\\
        E-mail: \email{rolf.walder@ens-lyon.fr}}

\abstract{We present the results from a particle-in-cell (PIC) simulation that models the interaction between a spatially localized electron-positron cloud and an electron-ion plasma. The latter is permeated by a magnetic field that is initially spatially uniform and aligned with the mean velocity vector of the pair cloud. The pair cloud expels the magnetic field and piles it up into an electromagnetic piston. Its electromagnetic field is strong enough to separate the pair cloud from the ambient plasma in the direction that is perpendicular to the cloud propagation direction. The piston propagates away from the spine of the injected pair cloud and it accelerates the protons to a high nonrelativistic speed. The accelerated protons form an outer cocoon that will eventually become separated from the unperturbed ambient plasma by a fast magnetosonic shock. No electromagnetic piston forms at the front of the cloud and a shock is mediated here by the filamentation instability. The final plasma distribution resembles that of a hydrodynamic jet. Collisionless plasma jets may form in the coronal plasma of accreting black holes and the interaction between the strong magnetic field of the piston and the hot pair cloud may contribute to radio emissions by such objects.}

\FullConference{High Energy Phenomena in Relativistic Outflows VII - HEPRO VII\\
		9-12 July 2019\\
		Facultat de Física, Universitat de Barcelona, Spain}

\begin{document}

\section{Introduction}

Accreting black holes in X-ray binaries, which are also known as microquasars, can emit jets that are composed of electrons, positrons and ions. Collimated jets of microquasars can move at relativistic bulk speeds \cite{Mirabel94}. Their source is thought to be the disk corona. It surrounds the inner accretion disk of the black hole and is fed by disk material that has been evaporated for example by viscous dissipation \cite{King07} and magneto-rotational instabilities \cite{Hawley95,Inchingolo18}. 

Thermal emissions in the soft X-ray range by microquasars suggest that the temperature of the inner disk is about 1000 electronVolts (1 keV). Hard X-ray radiation is attributed to its corona. Reconnection \cite{DalPino05} of magnetic field lines of the disk has been proposed as a mechanism that could transfer energy from the disk to the corona, which would maintain its high temperature. 

The spectral distribution of the radiation, which is emitted by the disk and its corona, depends on the state this system is in \cite{Remillard06}. Thermal radiation accounts for most of the electromagnetic emissions during the high-soft state and the cut-off energy of the nonthermal radiation is about 100 keV. Non-thermal emissions dominate in the low-hard state and their cut-off energy is of the order of 200 keV.  Relativistic jets are launched during the low-hard state \cite{Corbel00}. We refer to \cite{Yuan14} for a review of the physics of accretion disks and their coronae.

Observations of pair annihilation lines during flares of V404 Cygni \cite{Siegert16} demonstrate that processes close to the inner accretion disk of a black hole can create large clouds of electrons and positrons. Collisions between photons can produce pairs provided that their energy is high enough \cite{Guilbert83}. Photons with energies of the order of 200 keV, which are emitted by X-ray binaries during their low-hard state, do not carry enough energy. However, a local energy release, for example, by magnetic reconnection or by a shock wave could heat up the plasma and extend the associated photon spectrum beyond the energy that is required to create pairs. Pair creation by photon-photon interaction is efficient for the conditions found close to accretion disks. It results in a dense population of mildly relativistic electron-positron pairs and it absorbs photons with energies exceeding one MeV, which may explain why they are not emitted during flares. 

It is likely that these pair clouds form in the corona, which is a collisionless plasma that consists of electrons and ions. The high thermal velocity of electrons and positrons, their interaction with the background magnetic field and with the electromagnetic radiation that is emanated by the inner disk is likely to create pair outflows that move at least at mildly relativistic speeds. 

How does such an outflow interact with the ambient material? Contact discontinuities keep both fluids separated in hydrodynamic jet models \cite{Marti97,Bromberg11}. Such discontinuities form easily in a collisional fluid. Their thickness is comparable to the mean free path of its particles, which is likely to be long in the dilute coronal plasma. Furthermore, the low particle collision frequency implies that hydrodynamic contact discontinuities may not be established quickly. Discontinuities, which are sustained by the collective electromagnetic field of the plasma, may form faster than their hydrodynamic counterparts. Their properties, like their magnetic field amplitude and their ability to block a relativistic pair flow, determine the structure of a collisionless jet and its electromagnetic emission spectrum. It is thus important to explore these structures in more detail. 

We show here results from a particle-in-cell (PIC) simulation using the EPOCH code \cite{Arber15} where a collisionless discontinuity and the subsequent formation of a jet were observed \cite{DieckmannAA19}. The discontinuity was sustained by electric and magnetic fields and hence we refer to it as the electromagnetic piston. It is capable of separating the protons of the ambient plasma from the injected pair plasma. The expelled protons are accelerated to a high nonrelativistic speed. They will eventually drive a collisionless shock in the ambient plasma that will form the boundary of the jet's outer cocoon. The head of the jet is mediated by a filamentation instability between the ambient electrons and the pairs of the pair cloud \cite{Dieckmann18}. Its magnetic field moves relative to the protons, which accelerates them by their convective electric field \cite{Dieckmann18,Lemoine19}.  

Our paper is structured as follows. Section 2 summarizes the initial conditions of the simulation and presents its results. The latter are summarized in Section 3.

\section{Simulation}
 
Our two-dimensional simulation box is oriented in the x-y plane and filled uniformly with electrons with the mass $m_e$ and protons with the mass $m_p=1836m_e$. Both species of this ambient plasma have the number density $n_0$ and temperature $T_0$ = 2 keV. Space is normalized to the proton skin depth $\lambda_s = c/\omega_{pp}$, where $c$ is the speed of light and $\omega_{pp}={(n_0e^2/m_p\epsilon_0)}^{1/2}$ ($e$, $\epsilon_0$: elementary charge and dielectric permittivity). The electron plasma frequency is $\omega_{pe}={(n_0e^2/m_e\epsilon_0)}^{1/2}$. We resolve the interval $0 \le x \le L_x$ with the size $L_x= 24.6$ along $x$ by 14000 cells and the interval $-L_y/2 \le y \le L_y/2$ with the length $L_y= 12.3$ along $y$ by 7000 grid cells. Both species are resolved by 14 computational particles (CPs) each. A magnetic field $\mathbf{B}_0=(B_0,0,0)$ with the electron gyro-frequency $\omega_{ce}=eB_0/m_e = \omega_{pe}/11.3$ permeates the plasma at the starting time $t=0$. It has the normalized magnetic pressure $P_{B0}=\mathbf{B}_0^2/2\mu_0P_0$ ($\mu_0$: vacuum permeability) with the thermal pressure $P_{0}=n_0k_BT_0$ ($k_B$: Boltzmann constant). 

A pair cloud is injected at $x=0$. The cross-section of the density distribution of the injected electrons is $n_-(y)=5(1-{[y/1.7]}^2)$ if $n_-(y) \le 0$ and $n_-(y)=0$ otherwise. We use the same distribution $n_+(y)=n_-(y)$ for the positrons. We refer to these particles as the cloud particles. We set the temperature and mean speed of the cloud particles to 400 keV and to $v_b=0.9c$ along $x$. A total of $\approx 6 \cdot 10^5$ CPs are injected at every time step in order to resolve the cloud plasma. The simulation time $t_{sim}=33.75$ (unit $\omega_{pp}^{-1}$) is resolved by 61500 time steps.

We motivate our choice for the plasma temperatures as follows. The exact conditions of the coronal plasma are not known. The temperature of our ambient plasma is higher than that of the inner accretion disk, as expected for the coronal plasma. It is cooler than the nonthermal X-ray radiation of microquasars suggests. Our value $T_0=$ 2 keV should thus be a lower bound for the coronal temperature. Forthcoming simulation studies will take into account higher temperatures of the ambient plasma and allow us to identify how thermal effects affect the jet structure. 

The temperature and mean speed of the pair cloud should be representative for a pair outflow far from its source but still close to the microquasar. We do not resolve the source region itself because the temperature $T_0$ of the ambient plasma is too low for pair creation. The source of the outflow is likely to be a localized pair cloud with a mildly relativistic temperature \cite{Guilbert83}. The mean speed and temperature of the pair outflow should be comparable to the thermal speed of the source plasma. In time and in the absence of radiation and plasma instabilities, the pairs at the front of the cloud will cool down because the front contains only the fastest particles of the initial outflow. However, interactions of the cloud particles with the coronal X-ray radiation and plasma instabilities will maintain a high temperature of the pair cloud. Relevant instabilities are discussed in Ref. \cite{Nishikawa16} and the PIC simulations demonstrate how they heat up the cloud particles. We also point out that we are studying the discontinuity between the ambient plasma and the plasma of a jet; the pair plasma has been heated in this case by internal shocks near this discontinuity.

In what follows, we show the simulation results at the time $t_{sim}$ and in the half-plane $y<0$, where $y=0$ coincides with the central axis of the injected pair cloud. The plasma state prior to this time and for all other values $-L_y/2 \le y \le L_y/2$ is discussed in Ref. \cite{DieckmannAA19}. 

\begin{figure*}
\includegraphics[width=\textwidth]{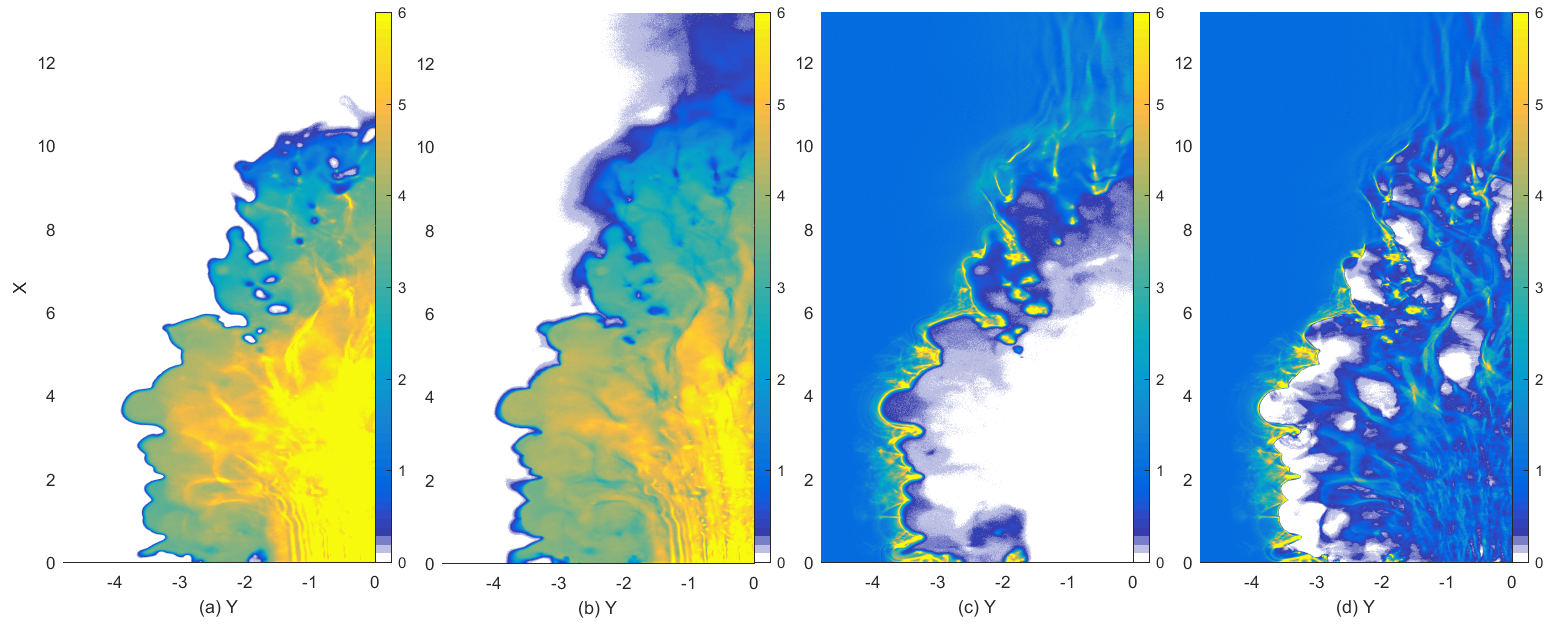}
\caption{Plasma density distributions at the time $t_{sim}$: Panels (a) and (b) show the densities of the cloud electrons and positrons. Panel (c) depicts the density of the ambient electrons and (d) that of the protons. All densities are normalized to $n_0$.}
\label{figure1}
\end{figure*}

Figure \ref{figure1} depicts the density distributions of the positrons, protons, cloud electrons and ambient electrons.
Figures \ref{figure1}(a,b) reveal a dense beam of cloud particles in the interval $x < 1.5$ and $y>-1.7$; these are the electrons and positrons we inject at the boundary $x=0$. This beam widens at larger values of $x$. The cloud maintains a high density up to $x\approx 7.5$. The cloud is confined to $y>-4$ in the interval $x<6$ and its width along $y$ decreases with increasing $x>6$. Figure \ref{figure1}(b) reveals a dilute positronic outflow at $x>11$ with no matching electronic counterpart. 

Figure \ref{figure1}(c) demonstrates that the ambient electrons have been expelled from the interval where the density of the cloud electrons exceeds the value 5. Their density peaks at the outer boundary of the cloud electrons reaching values above $6$ for $x<8$ and $y\approx -3.5$. The density distribution of the ambient electrons is unperturbed for larger distances from where the cloud particles are injected except for the interval $x>9$ and $y>-1.5$, where it has density striations that are are aligned with the x-axis. Similar striations are seen in the proton density in Fig. \ref{figure1}(d). These density striations arise from a filamentation instability between the outflowing positrons and the ambient plasma.
   
The proton density shows density filaments also in the interval $y>-3$ and $x<8$. They are the product of a filamentation instability between the ambient plasma and the fast-moving pair cloud, which developed at an earlier time. Proton filaments have much larger density gradients in the interval $x<8$ than those in the interval $x>10$. The filamentation instability at low $x$ is driven by a much denser pair cloud and, hence, it has much more energy available to compress the protons. Figure \ref{figure1}(d) reveals a pronounced gap for $-3.6 \le y \le -3$ and $x < 5.5$ between the filamentary structures at larger $y$ and the unperturbed ambient plasma at lower values of $y$. This gap is wider than the separation between the proton filaments close to the location where the pair cloud is injected and it must be sustained by a separate mechanism.   

\begin{figure*}
\includegraphics[width=\textwidth]{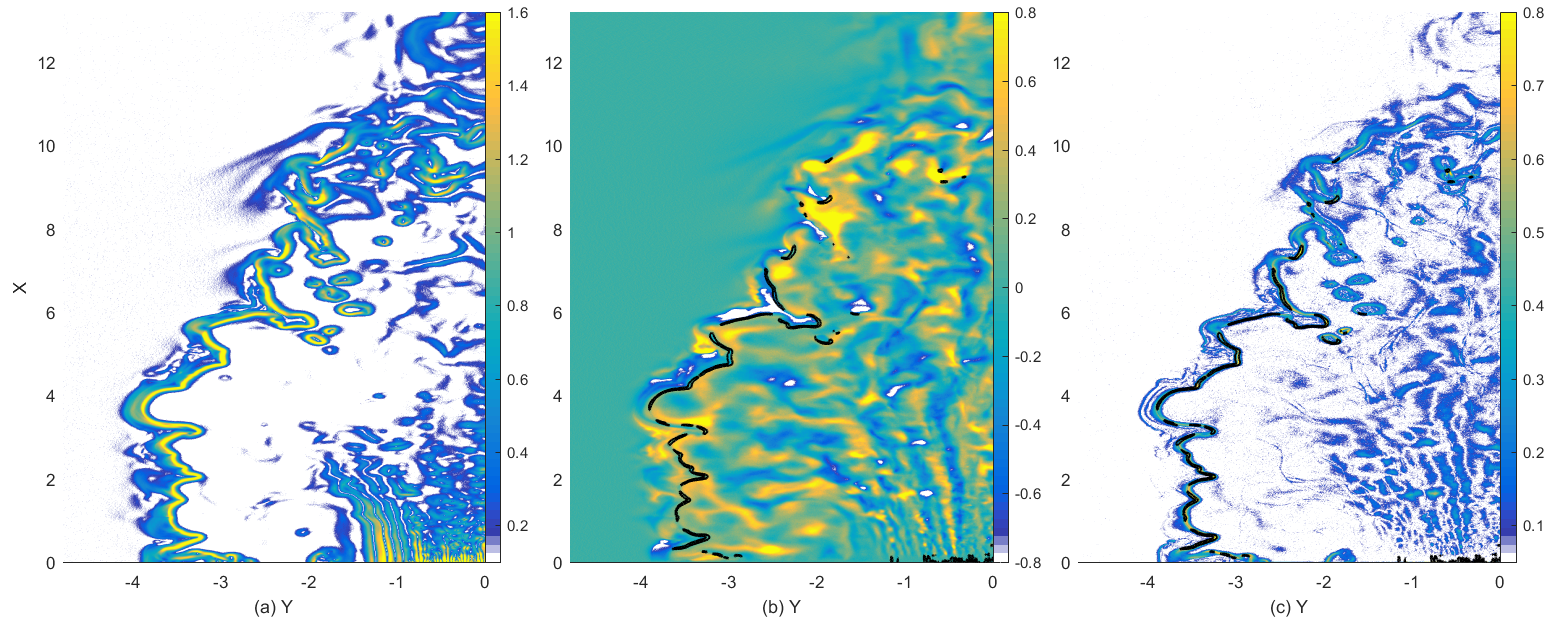}
\caption{The electromagnetic fields at the time $t_{sim}$: Panels (a) and (b) show the in-plane magnetic field $B_p = {(B_x^2 + B_y^2)}^{1/2}$ and $B_z$, respectively. Both are normalized to $m_e \omega_{pe}/e$. Panel (c) shows the in-plane electric field $E_p={(E_x^2+E_y^2)}^{1/2}$, which is normalized to $m_e c \omega_{pe}/e$. The black curves in (b, c) mark the contour line $B_p(x,y)=2$.}
\label{figure2}
\end{figure*}

Figure \ref{figure2} depicts the electromagnetic field components that are associated with the plasma density distributions. Magnetic and electric field amplitudes are normalized to $m_e\omega_{pe} / e$ and $m_e c \omega_{pe}/e$.
A magnetic band with an amplitude $B_p \ge 1.6$ is observed for $x<8$. It coincides with the outer boundary of the pair cloud in Figs. \ref{figure1}(a, b) and of the interval from which the protons are expelled in Fig. \ref{figure1}(d). This magnetic band was created when the background magnetic field $\mathbf{B}_0$ was piled up into an electromagnetic piston by the expanding pair plasma \cite{DieckmannAA19}. 

Figure \ref{figure2}(b) reveals modulations of $B_z$ close to the magnetic band and in the interval enclosed by it. A magnetic field $B_z\neq 0$ requires a current that flows in the simulation plane. The electrons and positrons of the cloud are forced apart by the presence of proton density modulations and the need to maintain a quasi-neutrality of the plasma. Their currents do not cancel each other out any more and the ensuing net current sustains the out-of-plane magnetic field. An in-plane electric field $E_p$ follows $B_p$ closely in Fig. \ref{figure2}(c). It points away from the cloud particles. 

\begin{figure*}
\includegraphics[width=0.5\columnwidth]{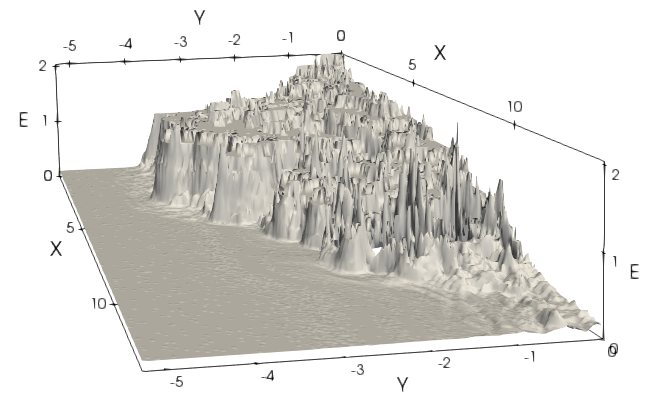}
\includegraphics[width=0.5\columnwidth]{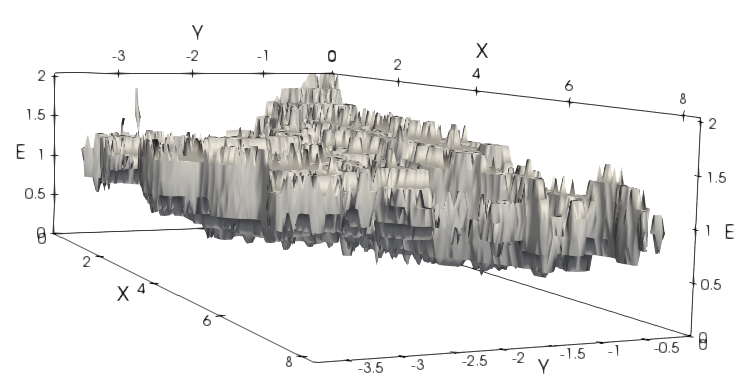}

\includegraphics[width=0.5\columnwidth]{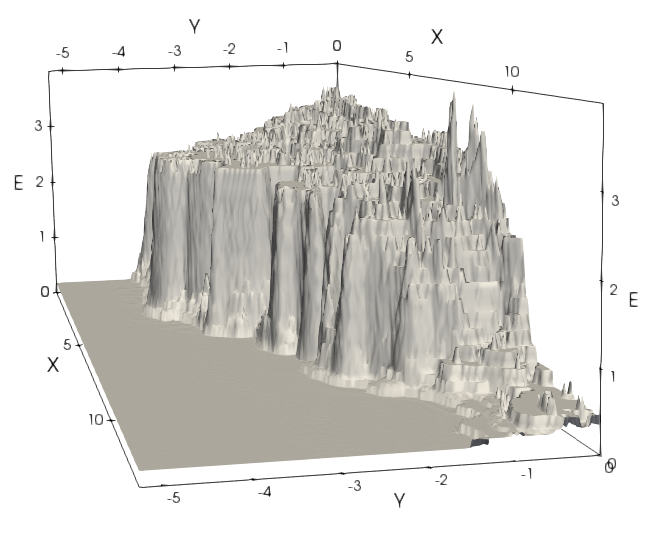}
\includegraphics[width=0.5\columnwidth]{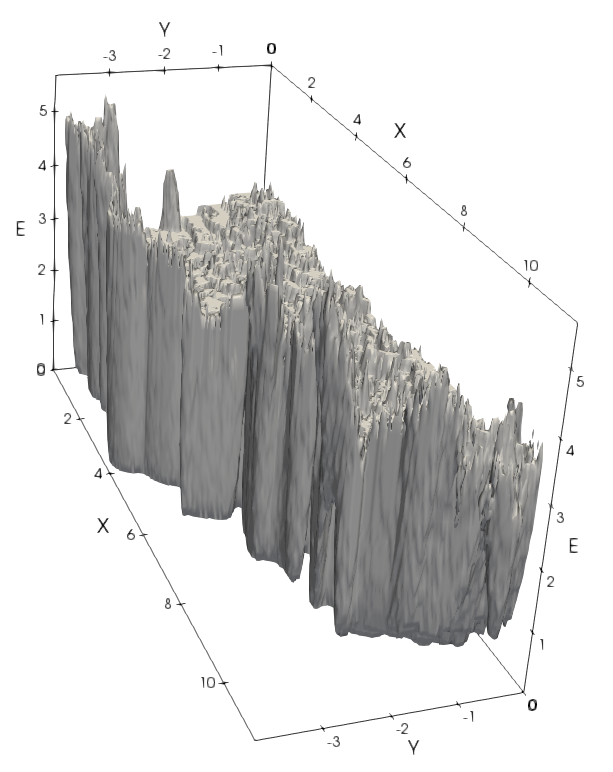}
\caption{Isosurfaces of the lepton distributions at the time $t=t_{sim}$, which are normalized to the peak density of the electron distribution: The upper row shows the phase space density contour $f_e(x,y,E)=0.05$ of the electrons (left) and $f_p(x,y,E)=0.05$ of the positrons (right). The lower row shows $f_e(x,y,E)=0.005$ (left) and $f_p(x,y,E)=0.005$ (right). Energies $E$ are expressed in units of $m_ec^2$.}
\label{figure3}
\end{figure*}

Figure \ref{figure3} plots two iso-contours of the phase space density distributions of positrons $f_p (x,y,E)$ and electrons $f_e(x,y,E)$ where the kinetic energy $E$ is expressed in units of the lepton's rest mass energy. We have summed up the distributions of both electron species. Both phase space density distributions are normalized to the peak value of the electron phase space density.
The shape marked by the contour 0.05 encloses the dense core distributions of both species. Cloud particles are fastest at low positions $x,|y|$. These positions correspond to the location where cloud particles are injected. The positrons are arranged in the form of a hot beam centered around $E\approx 1$. Positrons move at a mildly relativistic speed while the jet expands along $x$ at the nonrelativistic speed 0.15 $c$ \cite{DieckmannAA19}. Injected positrons are thus bound to encounter the jet's head and be reflected by it. This reflection is typically elastic. Heated particles spread in phase space, which results in a decrease of their phase space density. The isosurface 0.05 does not capture the heated population. 

The isosurface 0.05 of the electron distribution shows a well-defined front between the energetic electrons of the jet and the ambient electrons at rest. The distributions of the cloud and ambient electrons are separated by a small gap on the sides of the jet (not shown), which coincides with the location of the magnetic band in Fig. \ref{figure2}(a). The thermal gyro-radius of the ambient electrons is an order of magnitude smaller than the width of the magnetic band and hence they cannot cross it. They are pushed away from the jet axis by the electromagnetic piston that moves to increasing values of $|y|$. The speed of the piston is nonrelativistic and its expansion cannot accelerate the ambient electrons to relativistic speeds. Electrons with relativistic energies thus pertain to the cloud. The distributions of the ambient and cloud electrons mix at large $x$ and small $|y|$ because the magnetic band is not a closed structure for $x\ge 8$ and $|y|<2$.

Figure \ref{figure3} has demonstrated that the contours $f_{e,p}(x,y,E)=0.05$ reach about the same peak energy. The bulk of the cloud particles is thus moving at the same speed. This is not the case for the contours $f_{e,p}=0.005$ that are also shown in Fig. \ref{figure3}. Both contours differ strongly at the jet boundary. Electrons are decelerated at the magnetic band while positrons gain energy. The reason for this discrepancy is the electric field band in Fig. \ref{figure2}(c), which follows the magnetic band. The electric field is driven by the current of the expelled ambient electrons. It accelerates the protons away the cloud's interior, which results in the wide band with no protons in Fig. \ref{figure1}(d). Positrons are accelerated by the electric field as they approach the magnetic field band while cloud electrons are decelerated.

\begin{figure}
\includegraphics[width=\columnwidth]{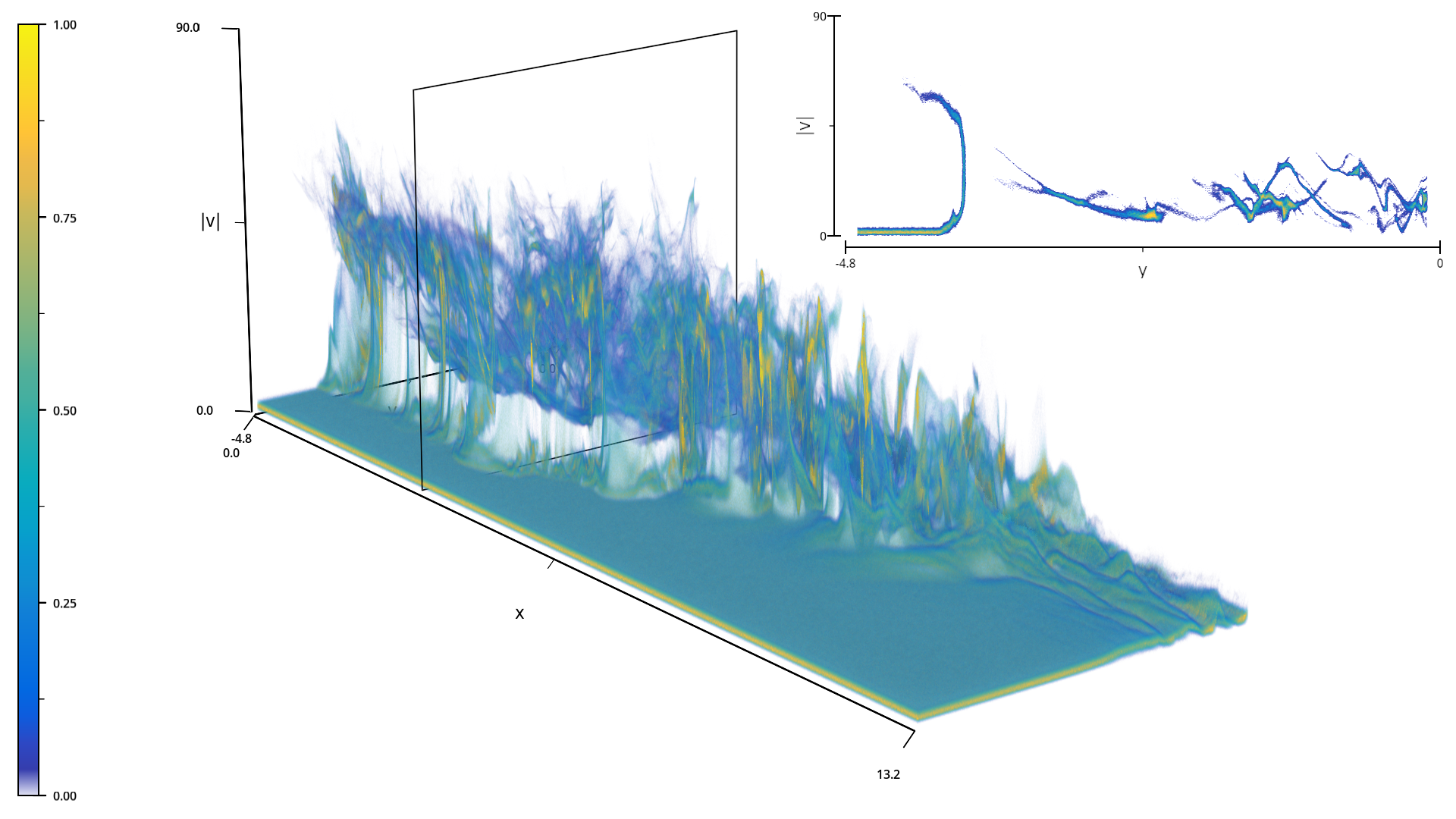}
\caption{Proton phase space density distribution $f_i(x,y,|v|)$ at the time $t_{sim}$. Values of $y$ increase as we go to the right and the right boundary corresponds to the symmetry axis $y=0$ of the injected pair cloud. The color scale is linear. The inset on the upper right slices the phase space density distribution at $x=3.5$ (black square). The rendering was created with Inviwo \cite{Jonsson19}.}
\label{figure4}
\end{figure}

Figure \ref{figure4} illustrates in more detail the effect the electric field band has on the protons; their rotation in the magnetic field is negligible. Proton velocities are normalized to their initial thermal speed $v_{tp}={(k_BT_0/m_p)}^{1/2}$.
The electromagnetic piston accelerates the ambient protons to $70 v_{tp}$ over a distance that is much less than $\lambda_s$. The acceleration is most effective at $x<7$ where this electromagnetic structure expands laterally to the jet spine. The inset in Fig. \ref{figure4} slices the proton distribution along the axis $x=3.5$. Protons are accelerated at $y\approx -4$. They are reflected almost specularly by the moving electromagnetic piston and a beam is created that moves at the speed $|v|\approx 60v_{tp}$ to decreasing $y$. This speed corresponds to about $0.1c$. A fast magnetosonic shock, which separates the unperturbed ambient plasma from the outer cocoon of the jet, will grow in time at the front of the reflected proton beam. 

\section{Summary}

We have examined the expansion of a hot cloud, which was composed of electrons and positrons with the same density distribution and velocity profile, into an ambient plasma that consisted of electrons and protons. A background magnetic field was aligned with the mean velocity vector of the pair cloud. The pair cloud had a limited spatial extent in the lateral direction. The purpose of our study here and in Ref. \cite{DieckmannAA19} was to determine if the pair plasma thermalizes with the ambient plasma via collisionless instabilities or if such an expansion could give rise to novel plasma structures. 

Our simulation results demonstrate that the pair cloud expels the protons on its way with the help of an electromagnetic piston. This piston is the collisionless counterpart of the contact discontinuity that separates the jet material from the ambient material in hydrodynamic jet models \cite{Marti97,Bromberg11}. Ambient electrons cannot cross the magnetic field of this piston and they are expelled by it. Their current drives an electrostatic field strong enough to evacuate the protons from the jet channel. The accelerated protons will eventually drive a shock in the ambient material that will form the border of the jet's outer cocoon. No electromagnetic piston could form at the head of the jet due to the chosen magnetic field geometry; the external shock at the jet's head is mediated by the filamentation (beam-Weibel) instability. Our simulation study did not reveal an internal shock between the fast-moving jet material and the electromagnetic piston. Such a shock would generate the hot pair population next to the electromagnetic piston that is known as the inner cocoon. Such a shock did presumably not form because of the high temperature of our injected pair cloud.  An internal shock will only form if the relative speed between the jet material and that of the inner cocoon exceeds the acoustic speed in the pair plasma.\\

\textbf{Acknowledgements} 

MED acknowledges financial support by a visiting fellowship from the Ecole Nationale Supérieure de Lyon, Université de Lyon. DF and RW acknowledge support from the French National Program of High Energy (PNHE). The simulations were performed with the EPOCH code financed by the grant EP/P02212X/1 on resources provided by the French supercomputing facilities GENCI through the grant A0050406960.

\end{document}